# Atomically Resolved FeSe/SrTiO$_3$(001) Interface Structure by Scanning Transmission Electron Microscopy

Fangsen Li[1*], Qinghua Zhang[2*], Chenjia Tang[1], Chong Liu[1], Jinan Shi[3], CaiNa Nie[1], Guanyu Zhou[1], Zheng Li[1], Wenhao Zhang[1], Can-Li Song[1,4], Ke He[1,4], Shuaihua Ji[1,4], Shengbai Zhang[5], Lin Gu[3,4], Lili Wang[1,4], Xu-Cun Ma[1,4] and Qi-Kun Xue[1,4]

*1 State Key Laboratory of Low-Dimensional Quantum Physics, Department of Physics, Tsinghua University, Beijing 100084, China*

*2State Key Laboratory of New Ceramics and Fine Processing, School of Materials Science and Engineering, Tsinghua University, Beijing 100084, China*

*3 Institute of Physics, Chinese Academy of Science, Beijing 100190, China*

*4Collaborative Innovation Center of Quantum Matter, Beijing 100084, China*

*5Department of Physics, Applied Physics and Astronomy, Rensselaer Polytechnic Institute, Troy, New York 12180, USA*

[*]These authors contributed twork.

*Email: l.gu@iphy.ac.cn, liliwang@mail.tsinghua.edu.cn,*

Interface-enhanced high-temperature superconductivity in one unit-cell (UC) FeSe films on SrTiO$_3$(001) (STO) substrate has recently attracted much attention in condensed matter physics and material science. By combined *in-situ* scanning tunneling microscopy/spectroscopy (STM/STS) and *ex-situ* scanning transmission electron microscopy (STEM) studies, we report on atomically resolved structure including both lattice constants and actual atomic positions of the FeSe/STO interface under both non-superconducting and superconducting states. We observed TiO$_2$ double layers (DLs) and significant atomic displacements in the top two layers of STO, lattice compression of the Se-Fe-Se triple layer, and relative shift between bottom Se and topmost Ti atoms. By imaging the interface structures under various superconducting states, we unveil a close correlation between interface structure and superconductivity. Our atomic-scale identification of FeSe/STO interface structure provides useful information on investigating the pairing mechanism of this interface-enhanced high-temperature superconducting system.

**Introduction**

The discovery of a superconducting gap $\Delta$ ~20 meV, nearly one order of magnitude higher than that of bulk FeSe, in one unit-cell (UC) FeSe film grown on SrTiO$_3$(001) (STO) substrate has stimulated much attention in superconductivity community [1]. The superconducting transition temperatures ($T_c$) ranging from 50 K to 100 K were subsequently revealed by transport [2-5] and mutual inductance measurements [2, 6]. Compared with $T_c$ ~9 K for bulk FeSe [7], the findings indicate a vital role of STO substrate in boosting the high temperature superconductivity. Previous angle-resolved photoemission spectroscopy (ARPES) [8-10] and scanning tunneling microscopy/spectroscopy (STM/STS) studies [11] have revealed significant charge transfer from STO substrate to FeSe films, which can be controlled by vacuum annealing [9, 11]. It is widely believed that the electrons injected into FeSe films originate from oxygen vacancies in STO [10-13]. However, atomic structure of the interface and how the doping is realized with annealing are still elusive. In addition to interface charge transfer, strong coupling between FeSe electrons and high-energy optical phonons in STO is evidenced by ARPES as occurrence of a replica band separated by ~100 meV from the main band [14]. In terms of the energy of ~100 meV, the responsible phonon should be a polar phonon mode on the surface layer of the STO substrate that is related to Ti-O displacement near surface [14-17]. Very recently, features of electron-phonon coupling with two low energy phonon modes at ~11 meV and ~21 meV, are observed in STS studies [18], in good agreement with the first-principle calculation that takes the template effect of STO substrate into account [19]. To understand these phenomena and superconductivity mechanism, atomic-scale information on the FeSe/STO interface is indispensable. In this study, we combine high-resolution scanning transmission electron microscopy (STEM) and STM/STS to investigate both in-plane and out-of-plane atomic structure of the interface. By studying the interface structures under both non-superconducting and superconducting samples, we reveal a close relation between interface structure and superconducting properties.

**Results and discussion**

Figure 1 shows the evolution of surface morphology, corresponding tunneling spectra and interface structure of 1-UC FeSe films on STO with ultra-high vacuum (UHV) annealing. The atomically resolved STM images (Figs. 1(a)-(c)) all display Se-terminated (001) FeSe surface with an in-plane lattice constant of ~ 3.90 Å, which is the same as that of STO(001) surface and thus 3% larger than that of bulk FeSe [7]. Dumbbell-like features (labeled as μ) are observed in Fig. 1(a) and 1(b), which correspond to extra Se in Se-Fe-Se triple layers under the Se-rich growth condition [20]. When the extra Se desorb with annealing, the 1-UC FeSe films convert from non-superconducting (black curve in Fig. 1(d)) to superconducting state with a gap of 12 meV and 17 meV (red and blue curves in Fig. 1(d), respectively). To simplify the description, we named the three different states as non-superconducting (NS), intermediate superconducting (IS), and optimal superconducting (OS) states, respectively. For the three states, the FeSe/STO interfaces are atomically sharp, as demonstrated by high-angle annular dark-filed (HAADF) images viewed along the [010] direction (Fig. 1(e)-1(f)).

Displayed in Fig. 2 are detailed STEM results of the FeSe/STO interface in OS state. Fig. 2(a) and 2(c) (Fig. 2(b) and 2(d)) show typical atomically resolved HAADF (annular bright-field (ABF)) images viewed along the [010] and [$\bar{1}$10] directions, respectively. In the HAADF images, Se, Fe, Sr and Ti atoms appear as bright dots and are clearly identified with sharp intensity contrast, while in the ABF images those atoms appear as dark dots and O atoms bright. An ordered Se-Fe-Se triple layer is clearly observed, indicative of nearly stoichiometric high-quality FeSe films. On the STO side, $TiO_2$ and SrO layers stack in sequence. Intriguingly, an extra layer, marked by blue circles in Fig. 2(a)-(d), is noticed right below the bottom Se layer. The atom columns in this extra layer is located nearly at the positions of Sr columns, but exhibit intensity similar to Ti columns. With O atoms in this extra layer disclosed in the corresponding ABF images, marked by red circles in Fig. 2(b) and 2(d), we infer the extra layer as either SrO or $TiO_x$ layer, which could cause quite different interface properties, e.g. the case of $LaAlO_3$/$SrTiO_3$ interface [21].

To identify the chemical composition of the extra layer, we conducted atomically resolved electron energy loss spectroscopy (EELS) of Ti-$L_{23}$ edge across the FeSe/STO interface. The results are shown in Fig. 2(e). At points far away from the interface, such as point "3", the spectrum shows four peaks in an energy range of 460 eV ~ 470 eV, the typical characteristic of Ti-$L_{23}$-edge observed in bulk STO, indicative of $Ti^{4+}$ [22]. Near the interface (points 2, 1, 0 and -1), the four characteristic peaks decrease in intensity and merge into two wide humps, indicating that partial $Ti^{4+}$ atoms change to $Ti^{3+}$ as resulted from the oxygen vacancies near the surface of STO [22]. The features of $Ti^{3+}$ are clearly discerned at point "0" and "-1" just below the Se layer, explicitly demonstrating that the extra layer is $TiO_x$ layer. Thus, a $TiO_x$ double layers (DLs) at the FeSe/STO interface is identified. The fact that this $TiO_x$ DLs occurs at the FeSe/STO interface irrespective of NS, IS or OS states (Fig. 1(e)-(g)), together with the $TiO_x$ rich surface of STO(001) reported previously [23-25], indicates that the $TiO_x$ DLs is intrinsic characteristic of STO(001). Compared with previous STEM results [25], the $TiO_x$ DLs is much more clearly resolved because of the sharp interface between single crystalline FeSe films and STO. To distinguish the two $TiO_x$ layers, we name the extra $TiO_x$ layer as top $TiO_x$ layer, and the other as the 1st $TiO_2$ layer of bulk STO.

A closer examination of the $TiO_x$ DLs reveals some "tail frame"-like features, indicated by arrows in Figs. 2(a)-2(c), around Ti columns in the top $TiO_x$ layer. Comparing the HAADF images of FeSe/STO interfaces under various states shown in Figs. 1(e)-1(g), we find that the "tail frame"-like features decrease gradually in both density and weight from NS to IS and then OS state. Considering the corresponding surface morphology evolution, i.e. desorption of extra Se atoms in Se-Fe-Se triple layers with annealing, seen from the STM topography images shown in Figs. 1(a)-1(c), we argue that the "tail frame"-like features correspond to extra Se atoms at the FeSe/STO interface, which may incorporate into the top $TiO_x$ layer and compensate some oxygen vacancies on STO surface.

To bring more insight into the correlation between structure and superconductivity properties, we quantified the interface structure by measuring various inter-atomic distances at NS, IS and OS states depicted in the schematic model in Fig. 3(a). The results are summarized in Fig. 3(b) and Table 1. As indicated by Fig. 3(b), the distance between the double $TiO_x$ layers is 2.44 Å ±0.08 Å, which is ~25% larger than half of the lattice constant of STO along the [001] direction and independent of the sample states. Both the 1st $TiO_2$ and 1st SrO layer shrink inwards, as evidenced from the reduced nearest neighboring (NN) layer distance of 1.81 Å ±0.05 Å and 1.85 Å ±0.09 Å, respectively. For the 2nd $TiO_2$ and deeper layers, the atomic displacements are negligible. Consistent with the in-plane lattice of ~3.90 Å revealed by STM, identical in-plane lattice constant (the Se-Se distance) of ~3.86 Å is observed in the FeSe films under the three distinct states. Intriguingly, when the film changes from NS to OS state, the Se height $h_{Se}$ decreases from 1.33 Å ±0.02 Å to 1.31 Å ± 0.01 Å, whereas the separation $d_{Se-Ti}$ between the bottom Se layer and the topmost Ti layer increases from 3.34 Å ±0.05 Å to 3.58 Å ±0.05 Å (Table 1).

We now focus on the interface structure of the OS sample. The Se-Se distance is 3.86 Å ±0.03 Å, ~2.5% expanded compared to bulk FeSe [7]. To compensate the in-plane expansion, the Se height $h_{Se}$ reduces to 1.31 Å ±0.01 Å, ~9.5 % lower than that of bulk FeSe (1.45 Å) [7]. Consequently, the Se-Fe-Se bond angle α changes to ~ 111.4 ° ± 0.9 °, which is close to that of a regular tetrahedron (~109.5 °). Therefore, compared to bulk FeSe, the 1-UC FeSe films on STO exhibit more two-dimensional essential Se-Fe-Se triple layer (~9.5 % compressed $h_{Se}$) and much larger superconducting gap (~20 meV). This seems to be consistent with the empirical rule in bulk iron chalocogenides that the superconductivity is enhanced as the Se height $h_{Se}$ above the Fe layer is reduced [26, 27]. However, the value of $h_{Se}$ observed here is far from the optimal value of 1.38 Å [28]. Considering the previous observations that additional ~2% expansion in single-layer FeSe films [29] only induces a subtle increase of gap-closing temperature ($T_g$) of 5 K, and that anisotropic FeSe films on STO(110) substrates show comparable superconducting gap (17 meV) [30], we conclude that the tensile stress is not a key factor that induces the high temperature superconductivity in FeSe/STO. Consistent with the local distortion due to the lattice mismatch between FeSe and STO disclosed by atomically resolved STM images [18], the relative shift between the bottom Se atoms and the topmost Ti atoms along the in-plane direction is also clearly discerned from the STEM images in Fig. 2(f), which indicates the bonding between FeSe film and STO substrate can vary locally.

The atomic-resolved FeSe/STO interface structures shown above allow us to understand how the interface high temperature superconductivity is achieved by annealing. In the NS state, the extra Se atoms not only exist in FeSe films (dumbbell-like features in Fig.1(a)) but also incorporate in the top $TiO_x$ layer to compensate some oxygen vacancies at FeSe/STO interface ("tail frame"-like features in Fig. 1(e)). The extra Se atoms act as hole dopants [31, 32] and may partially or even totally counteract the electron doping from the oxygen vacancies. Correspondingly, NS samples exhibit both hole pocket slight below Fermi level at the Brillouin zone center and electron pockets at the zone corner [9]. Once these extra Se atoms in FeSe films and at FeSe/STO interface are removed after extensive annealing, oxygen vacancies are released, promoting charge transfer from STO to FeSe. The resulted electron doping pushes the hole bands at the Brillouin zone center from the Fermi energy [9]. Moreover, the phonon mode with energy ~80-100 meV, responsible for the interface enhanced electron-phonon coupling as revealed by ARPES study [14], corresponds to a special oxygen-vacancy induced flat phonon mode which is mainly composed of relative Ti and O atomic displacements along (001) direction in the top two layers and appear exclusively at the relaxed surface [17]. Accordingly, the release of oxygen vacancies with interfacial Se desorption can activate simultaneously the interface enhanced electron-phonon coupling. The inward shrinking of the 1st $TiO_2$ and 1st SrO layer (Fig. 3(b)) agrees qualitatively with the relaxed surface model proposed in Ref. [17]. This way, charge transfer and interface electron-phonon coupling could be boosted for the high temperature superconductivity in this system. The increase of ~ 0.24 Å in the interface separation $d_{Se-Ti}$ associated with the interfacial Se desorption (from NS to OS state) could be due to the formation of interface dipoles as resulted from charge accumulation at the interface [33-35].

## Conclusions

Atomically resolved STM and STEM images have been successfully achieved to reveal the atomic-scale interface structure of FeSe/STO. $TiO_x$ DLs with apparent atomic displacements are clearly identified at FeSe/STO interface. We observe a compressed Se-Fe-Se triple layer and a clear lattice shift between FeSe films and STO substrates. By studying the interface structures under different states, we unveil the correlation between interface structure and superconductivity properties. Our results reveal the important roles of charge transfer and interface electron-phonon coupling in the high temperature superconductivity in FeSe/STO systems.

## Materials and Methods

### Growth of FeSe films on $SrTiO_3$

One unit-cell (UC) FeSe films were epitaxially grown on Nb-doped $SrTiO_3$(001) (Nb: 0.5 wt%, KMT) substrates at 400 ℃ by co-depositing high purity Fe (99.995%) and Se (99.9999%) in standard Knudsen cells with a flux ratio of ~1:10. Prior to film growth, as-received Nb:STO substrates were treated by Se molecular beam etching method [1] to achieve atomically flat surface. The superconductivity state of FeSe films was controlled by annealing time [9] and annealing temperature [11]. To protect FeSe films from being oxidized for *ex-situ* STEM measurement, 10-

UC thick FeTe films were grown by co-evaporating Fe (99.995%) and Te (99.9999%) at a substrate temperature of 300 ℃.

**Scanning tunneling microscopy/spectroscopy (STM/STS) characterization**

STM/STS experiments were carried out in a low-temperature (down to 5.7 K) Createc STM system with a polycrystalline Pt/Ir tip. The STS tunneling spectra were acquired using lock-in technique with a bias modulation of 3 mV at 981.3 Hz. The STM images were processed using WSxM.

**Scanning transmission electron microscopy (STEM) characterization**

STEM experiments were performed on an aberration-corrected ARM-200CF (JEOL, Tokyo, Japan), which was operated at 200 keV and equipped with double spherical aberration (Cs) correctors. The attainable resolution of the probe defined by the objective pre-field is 78 picometers. For each EEL spectrum, the measured step is 0.2 nm. STEM samples were prepared using Focused Ion Beam (FIB) method. Cross-sectional lamella was thinned down to 100 nm thick at an accelerating voltage of 30 kV with a decreasing current from the maximum 2.5 nA, followed by fine polish at an accelerating voltage of 2 kV with a small current of 40 pA. All STEM images were corrected for drift with the lattice constant of Sr ~3.905 Å deep in the bulk. To precisely measure layers' distance along *c* direction, we projected the intensity of columns in HAADF images to [001] axis along [100] direction. Peaks in the projected curve correspond to the layers' vertical positions and the distance between adjacent peaks is defined as the vertical separation between layers.

## Acknowledgements

This work is supported by Ministry of Science and Technology of China (Grants No. 2015CB921000, No. 2014CB921002) and National Science Foundation of China (Grants No. 91421312, No. 91121004, No. 11574174) and the Strategic Priority Research Program of the Chinese Academy of Sciences (Grant No. XDB07030200). SBZ were supported by the US-DOE under Grant No. DESC0002623.

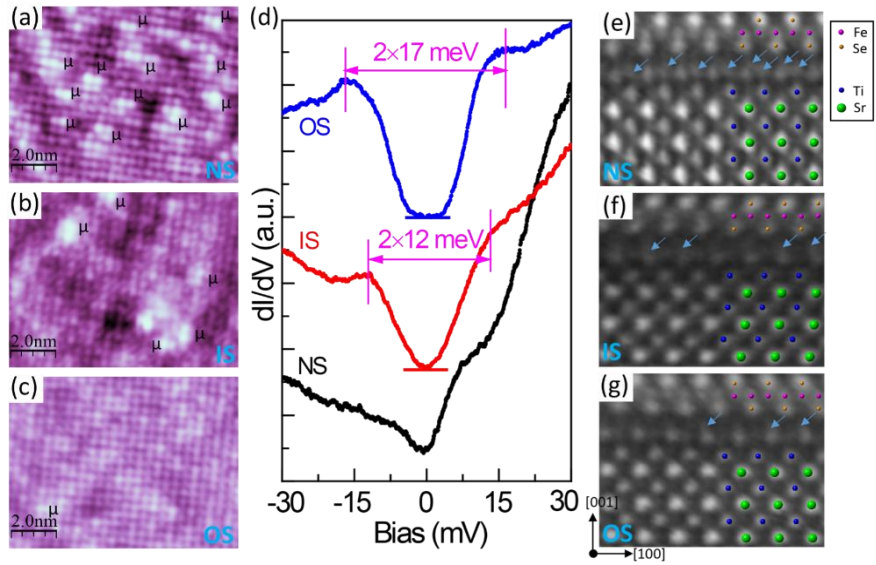

**Figure 1.** (a-c) STM images ($V = 1.0$ V, $I = 50$ pA, 7 nm×10 nm) of 1-UC FeSe on STO (001) substrates, showing evolution of surface morphology with vacuum annealing. The dumbbell-like features are labeled as µ. (d) Typical d$I$/d$V$ spectra taken on the surfaces shown in (a)-(c), showing the 1-UC FeSe films evolve from NS to IS and then OS states with increasing annealing. The d$I$/d$V$ curves of IS and OS states are shifted with the zero conductance indicated by horizontal bars. (e-g) HAADF images viewed along [010] direction of FeSe/STO under NS, IS and OS states, respectively. Arrows mark the "tail frame"-like features at the FeSe/STO interface.

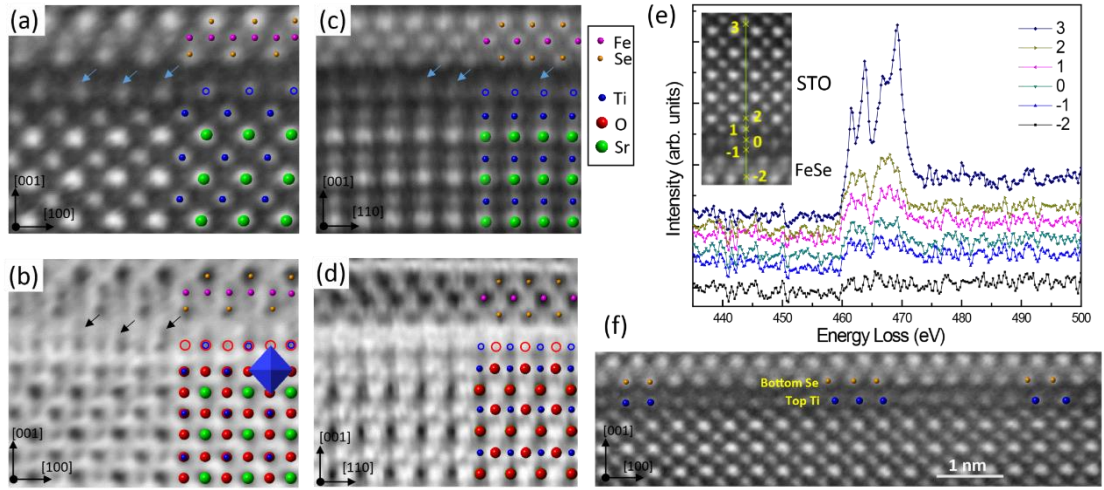

**Figure 2.** Atomically resolved STEM images of 1-UC FeSe/STO(001) heterostructure under OS state. (a) HAADF and (b) corresponding ABF images viewed along [010] direction. In (b), distorted oxygen octahedral was drawn. (c) HAADF and (d) corresponding ABF images viewed along [$\bar{1}$10] direction. The arrows in (a) and (b) marked the "tail-fame"-like features at FeSe/STO interface. (e) Local electron energy loss spectroscopy (EELS) of Ti-$L_{23}$-edge across FeSe/SrTiO$_3$ interface. Inset: HAADF image, showing the sites where the spectra were taken. (f) Large-scale HAADF image of FeSe/STO(001) heterostructure, showing relative shift between bottom Se atoms and topmost Ti atoms.

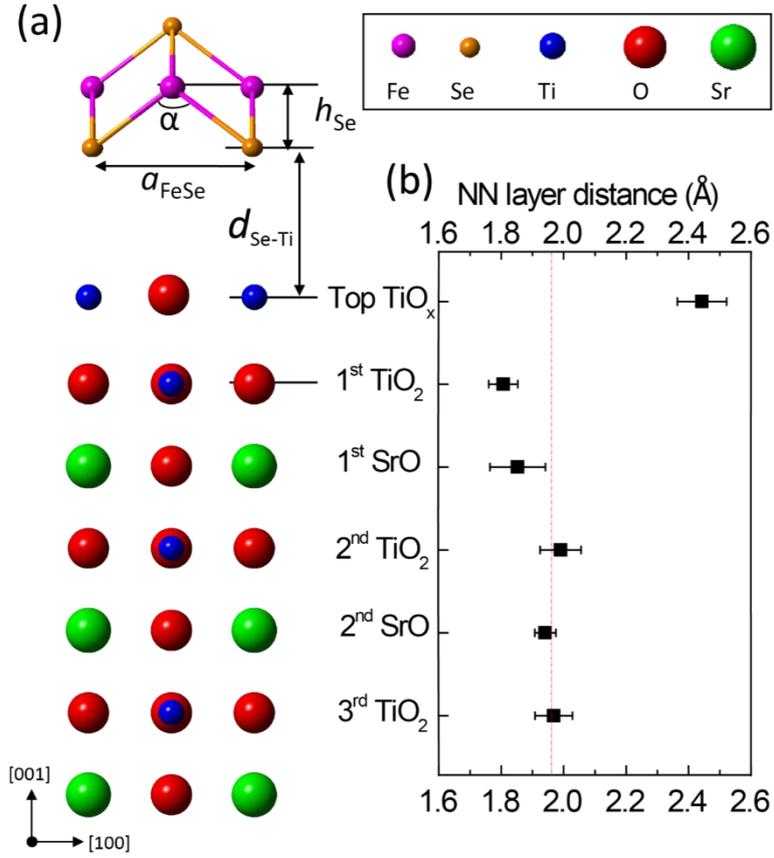

**Figure 3.** (a) Schematic model of FeSe/STO interface, showing 1-UC FeSe on TiO$_2$ DLs-terminated STO. (b) Nearest-neighboring (NN) layer distance along $c$ direction of STO, showing the displacement in the top two layers of STO surface.

**Table 1.** Structure parameters of FeSe/SrTiO$_3$ interface under various superconducting states (NS, IS and OS): lattice constant $a_{FeSe}$, anion height $h_{Se}$, angle $\alpha$ of Se-Fe-Se bonding, and distance $d_{Se-Ti}$ between bottom Se and topmost Ti atoms.

|  | NS | IS | OS |
|---|---|---|---|
| $a_{FeSe}$/Å | 3.87 ± 0.02 | 3.85 ± 0.02 | 3.86 ± 0.03 |
| $h_{Se}$/Å | 1.33 ± 0.02 | 1.32 ± 0.02 | 1.31 ± 0.01 |
| $\alpha$ | 111.0° ± 0.9° | 111.0° ± 0.7° | 111.4° ± 0.9° |
| $d_{Se-Ti}$/Å | 3.34 ± 0.05 | 3.47 ± 0.07 | 3.58 ± 0.05 |